\documentclass[12pt]{article}
\usepackage{a4}
\usepackage{bbold}

\def\bea{\begin{eqnarray}}
\def\eea{\end{eqnarray}}
\def\be{\begin{equation}}
\def\ee{\end{equation}}
\def\non{\nonumber \\}

\def\vs{v\!\!\! /}

\def\nn{\nonumber}

\newcommand\eqn[1]{(\ref{#1})}
\newcommand{\ft}[2]{{\textstyle\frac{#1}{#2}}}

\def\lra{\longrightarrow}

\newcommand{\Lap}{\bigtriangleup}
\renewcommand{\a}{\alpha}
\renewcommand{\b}{\beta}

\renewcommand{\d}{\delta}

\newcommand{\g}{\gamma} 
\newcommand{\e}{\epsilon}

\newcommand{\p}{\partial}

\renewcommand{\th}{\theta}

\newcommand\text[1]{\rm #1}

\begin{document}

\thispagestyle{empty}
\begin{flushright}
{\sc\footnotesize hep-th/0001106}\\[2mm]
{\sc AEI}-2000-002
\end{flushright}
\vspace{1cm}
\setcounter{footnote}{0}
\begin{center}
{\Large{\bf On the supersymmetric effective action of Matrix theory}
    }\\[14mm]
{\sc Hermann Nicolai and
Jan Plefka \\[7mm]
Max-Planck-Institut f\"ur Gravitationsphysik\\
Albert-Einstein-Institut\\
Am M\"uhlenberg 1, D-14476 Golm, Germany}\\[20mm]

{\sc Abstract}\\[2mm]
\end{center}
We present a simple derivation of the supersymmetric one-loop
effective action of SU(2) Matrix theory by expressing it in
a compact exponential form whose invariance
under supersymmetry transformations is obvious.
This result clarifies the one-loop exactness of the 
leading $(v^2)^2$ interactions and the absence of non-perturbative
corrections.

\vfill
\leftline{{\sc January 2000}}

\newpage
\setcounter{page}{1}

Recently maximally supersymmetric $SU(N)$ gauge quantum mechanics 
in $d=9$ \cite{CH} has gained prominence due to its relation to the 
low-energy dynamics of zerobranes in type IIA string theory \cite{d0}, 
the close relation between its $N\rightarrow\infty$ limit and the
eleven-dimensional supermembrane \cite{DWHN}, as well as the M theory 
proposal of \cite{BFSS}. A key feature of this model is the existence 
of flat directions in the Cartan sector on which scattering states 
localize. To date almost all investigations of scattering amplitudes 
in Matrix theory make use of the perturbative construction of an 
effective Lagrangian for the Cartan valley degrees of freedom at
finite $N$, which is based on a loopwise expansion around the solution 
of the classical equations of motions, $\ddot x^i_I=0$ and $\dot\th_I=0$,
where $I,J, \dots=1,...,N-1$. Although this approach simply ignores 
contributions from bound states {\it all} tree level amplitudes 
computed to date in matrix theory agree with the results obtained 
from eleven dimensional supergravity  \cite{treelevel,PSW}. As soon 
as one goes beyond the tree level regime, however, this correspondence
breaks down\cite{R4}\footnote{In fact, it can be shown that there is 
{\it no} Lorentz invariant combination of $R^4$ terms that reproduces 
the matrix theory result.}. As argued in \cite{PSS98,SSmore} the 
agreement of the $(v^2)^2$ and $(v^2)^3$ terms in the effective action
with tree level supergravity could be solely due to the high amount of
supersymmetry in the problem. In particular in \cite{PSS98} 
it was shown that the leading corrections to the $SU(2)$ effective 
action of order $(v^2)^2$ are completely determined by supersymmetry, 
a claim thereafter made explicit by \cite{HKS}. In this note we wish 
to demonstrate how this rather involved analysis may be
condensed to a two line argument, yielding the complete form of the 
supersymmetric $SU(2)$ effective action at order $(v^2)^2$.

The basic variables for the $SU(2)$ theory in the Cartan subalgebra are
\be
X^i (t) \quad , \quad v^i (t) := \dot X^i (t)  \qquad {\rm and} \qquad
\th_\a (t)
\ee
where $i,j,...=1,...,9$ and $\a,\b = 1,...,16$. These variables 
correspond to the diagonal degrees of freedom in the matrix 
theory; for $SU(N)$ we would have $X^i_I$ with $I=1,\ldots,(N-1)$.

The supersymmetry variations are given by:
\be
\d X^i = -i \e \g^i \th + \e N^i \th \qquad \d\th = v^j \g_j \e + M\e
\ee
where $N^i$ and $M$ correspond to higher order modifications.
For $N^i=M=0$, these variations leave invariant the free action
\be
S^{(0)} = \int dt \left[ \frac12 v^2 + \frac i2\th \dot \th \right]
\label{S0}
\ee
Besides these terms, the effective action will contain an infinite
string of higher order corrections. Since the algebra closes only
on-shell, the supersymmetry variations must be modified accordingly
such that $N^i$ and $M$ will no longer vanish. In considering such 
corrections, one must also take into account nonlinear field redefinitions
\be
X^i \lra X'^i=X'^i(X,v,\th) \qquad \th_\alpha \lra 
  \th'_\alpha=\th'_\alpha (X,v,\th) 
\ee
Modifications of the supersymmetry variations induced by such
redefinitions do preserve the algebra, but should be discarded
as they do not correspond to genuine deformations of the original
variations.

Remarkably, even in this simple quantum mechanical context, no 
nontrivial modifications with $N^i, M \neq 0$ have so far been 
explicitly exhibited in the literature, although in \cite{SSmore} 
evidence for non-trivial $N^i\sim\th^{4}$ and $M\sim\th^{6}$ 
modifications was presented. 
A full treatment is difficult because a complete analysis of
the superalgebra and its closure will presumably require 
the consideration of infinitely many corrections. The problem is 
aggravated by the fact that for the maximally extended models no 
off-shell formulation is known\footnote{Besides, it is doubtful
whether an off-shell formalism would be of much use 
here, as the ``rules of the game'' are no longer clear: the elimination 
of auxiliary fields via their equations of motion and via the path
integral yield inequivalent results unless the auxiliary 
fields appear at most quadratically in the Lagrangian.}.

To simplify matters, one makes the assumption that
\be
\frac{dv^i}{dt}= 0 \quad , \quad \frac{d\th}{dt} = 0
\label{constancy}
\ee
This assumption, which implicitly also underlies all previous work, 
allows us to drop all derivatives other than those of $X^i$
in the variations, and greatly simplifies the calculation; for 
instance, we can then consistently set $\d v^i =0$ for the unmodified
variations. Effectively, the above condition amounts to a reduction of
a quantum mechanical system to a ``zero-dimensional'' system.
The freedom of making field redefinitions is reduced accordingly:
the only redefinitions compatible with the above reduction are the
ones preserving the linear dependence of $X^i$ on $t$ and the
constancy of $v^i$ and $\th$.

In \cite{oneloop,PSW,matrixoneloop,HKS} the full supersymmetric one-loop 
effective action was shown to be 
\bea
S^{(1)}&=&\int dt \Big[ (v^2)^2 f(X) +\frac{i}2 v^2 v^j\p_j f\,(\th\g^{ij}\th) 
   -\frac18 v^i v^j \p_k \p_l f(X) \, (\th\g^{ik}\th) (\th \g^{jl} \th) \non
 && -\frac{i}{144} v^i\p_j \p_k \p_l f(X) \, (\th\g^{ij}\th) (\th\g^{km}\th)
         (\th\g^{lm}\th) \non
 && + \frac1{8064}\p_i \p_j \p_k \p_l f(X) \, (\th\g^{im}\th)(\th \g^{jm}\th)
         (\th \g^{kn}\th) (\th \g^{ln} \th) \Big]
\label{S1}
\eea
Here $f=f(X)$ is a harmonic function, i.e. $\Lap f \equiv \p_j\p_j f(X) =0$ 
with the unique rotationally invariant solution $f=r^{-7}$ (where
$r:=\sqrt{X^i X^i}$). Provided one assumes constancy of
$v^i$ and $\th$ the action $S^{(1)}$ must be invariant under the 
{\it unmodified} supersymmetry variations above, as the modified variation 
of the free action
$S^{(0)}$ under constant $v^i$ and $\th$
\be
\d S^{(0)}=\int dt\, \partial_t\, ( v^i\, \epsilon N^i\th)
\ee
vanishes for asymptotically suppressed corrections,  
$\lim_{t\rightarrow\pm\infty}N^i=0$.
Possible leading modifications of the supersymmetry transformations
were discussed in \cite{HKS}, but clearly these 
do not contribute under the assumption \eqn{constancy}.
                      
We will now show that the complicated action $S^{(1)}$ can be cast into a 
much simpler form, whose supersymmetry invariance is very easy to check.
Namely, we have 
\bea
S^{(1)} &=& \int dt\, 
(v^2)^2 \exp \left[ \frac{i}{2v^2} \th \g^{ij} \th v_i \p_j \right] f(X) \non
&=& \int dt\, (v^2)^2 f\left(X - \frac{i}{2v^2} \th \g^{ij} \th v_j \right)
\label{exp}
\eea
To prove that this action indeed coincides with \eqn{S1}, we first 
observe that in the above action we can neglect all terms containing 
the Laplacian (which annihilates $f$) as well as terms containing $v^j\p_j$, 
because with constant $v^i$ and $\th$, this term can be pulled out, 
yielding a total time derivative. To proceed, we show that
\be
(\th \g^{ij}\th v_i \p_j) (\th \g^{kl}\th \p_l) (\th \g^{km}\th \p_m)
    \simeq \frac3{v^2} (\th \g^{ij} \th v_i \p_j)^3  
\label{o3}
\ee
and
\be
\Big( (\th \g^{kl}\th \p_l) \, (\th \g^{km}\th \p_m) \Big)^2
  \simeq \frac{21}{(v^2)^2} \, \big(\th \g^{ij} \th v_i \p_j \big)^4 
\label{o4}
\ee
from which the equivalence follows up to fourth order. The symbol 
$\simeq$ here and below means equality modulo contributions containing
$v^i \p_i$ or $\Lap$. To verify the above relations we start out from the
Fierz identity (see e.g. \cite{HKS} for a comprehensive list
of Fierz identites)
\be
(\th \g^{ijk}\th \, v_i \p_j)^2 \simeq
      - 5 \, (\th \g^{ij} \th \, v_i \p_j )^2 + 
      v^2 \, (\th \g^{ij}\th \p_j) \, (\th \g^{ik}\th \p_k)
\label{1}
\ee
Thereafter one multiplies \eqn{1}
with $(\th \g^{ij}\th\, v_i \p_j)$ so that its 
left hand side may be rewritten as
\be
(\th \g^{ij}\th\, v_i \p_j)\, (\th \g^{klm}\th \, v_k \p_l)^2 \simeq
-\ft 1 6 v^2\, (\th \g^{ij}\th \p_i)\, (\th \g^{klj}\th\, \p_k)\,
(\th \g^{mnl}\th\, v_m\p_n) \, .
\ee
upon using yet another Fierz identity.
Now once more perform a Fierz rearrangement on the last two
terms of the above expression to obtain
\be
(\th \g^{ij}\th\, v_i \p_j)\, (\th \g^{klm}\th \, v_k \p_l)^2 \simeq
-\ft 2 3 v^2\, (\th \g^{ij}\th\, v_i \p_j)\, (\th\g^{kl}\th\, \p_k)^2
\label{a}
\ee
This is to be contrasted with the right hand side of \eqn{1} multiplied
with $(\th \g^{ij} \th \, v_i \p_j )$:
\be
(\th \g^{ij}\th\, v_i \p_j)\, (\th \g^{klm}\th \, v_k \p_l)^2
\simeq -5 \, (\th \g^{ij} \th \, v_i \p_j )^3 + 
      v^2 \, (\th \g^{ij} \th \, v_i \p_j )\, (\th \g^{kl}\th \p_k)^2
\label{b}
\ee
{}From \eqn{a} and \eqn{b} the relation \eqn{o3} immediately follows.
Relation \eqn{o4} is then shown in a similar manner.

Next we note that
the exponential series of \eqn{exp}
terminates already after the fourth order term because 
\be
(\th \g^{ij}\th v_i \p_j )^5 \simeq 0 \quad  ,
\label{o5}
\ee
which is an immediate consequence of \eqn{o3} and \eqn{o4}:
\bea
\frac{21}{(v^2)^2} \, \big(\th \g^{ij} \th v_i \p_j \big)^5&\simeq&
(\th \g^{ij} \th v_i \p_j \big)\,
\Big( (\th \g^{kl}\th \p_l) \, (\th \g^{km}\th \p_m) \Big)^2\nn\\
&\simeq&\frac3{v^2} (\th \g^{ij} \th v_i \p_j)^3  \, 
(\th \g^{kl}\th \p_l) \, (\th \g^{km}\th \p_m)\nn\\
&\simeq& \frac9{(v^2)^2} (\th \g^{ij} \th v_i \p_j)^5
\label{o5proof}
\eea
where we used \eqn{o4} in the first and \eqn{o3} in the second and
third lines.
Hence there is no need to truncate the series (it would anyhow terminate
at order $\th^{16}$ by the nilpotency of the Grassmann variables).

The supersymmetry of this action with the above assumptions \eqn{constancy}
(and $N^i = M = 0$) can now be proven in two lines:
\bea
\d S^{(1)} &=& \int dt (v^2)^2 
     \exp \left[ \frac{i}{2v^2} \th \g^{ij} \th v_i \p_j \right]
     \left( \frac{i}{v^2} \d \th \g^{ij}\th v_i \p_j f(X)
     + \d X^i \, \partial_i f(X) \right) \non
      &=& -\int dt (v^2)^2 
     \exp \left[ \frac{i}{2v^2} \th \g^{ij} \th v_i \p_j \right]
     \, \frac{i}{v^2} \,\epsilon\,\vs \,\th \, v^i\, \partial_i f(X) =0 \, .
\eea

Remarkably, this simple argument works {\it for any action} 
of the form
\be
\int dt \exp \left[ \frac{i}{2v^2} \th \g^{ij} \th v_i \p_j \right] g(X,v)
\label{anyaction}
\ee
and in particular yields supersymmetric completions of
\be
g(r,v)= \frac{(v^2)^m}{r^n}
\ee
Uniqueness and therefore a ``non-renormalization theorem'' holds only 
for actions with low powers of $v^2$, and only if one insists on the
absence of terms singular in $v^2$. For $g(r,v)\propto v^2$, the only
way to avoid such singular terms is to require $\p_i g =0$, in which
case one is left with the free action only. For $g(r,v)\propto (v^2)^2$, 
a singularity could arise at order $\th^6$, and is eliminated by means of 
the requirement $\Lap g = 0$ (and the above Fierz identities implying
\eqn{o5}).
Unfortunately for $g(r,v)\propto(v^2)^3$ our above arguments fail,
as the {\it modified} supersymmetry transformations of $S^{(1)}$
now do produce non-vanishing terms under the constancy assumption
of $v^i$ and $\th$. Hence even in this reduced sector \eqn{anyaction} 
cannot be the full answer. At order $g(r,v)\propto(v^2)^4$ and beyond, no
singular terms can arise, and the choice of $g(r,v)$ is not 
restricted in any way. This is meant by the statement that, at this 
order there is no ``non-renormalization theorem''.

We have thus seen that the one-loop effective action \eqn{S1}
resp.\ \eqn{exp} laboriously computed in \cite{oneloop,PSW,matrixoneloop,HKS}
is indeed completely fixed by supersymmetry. Therefore the agreement of
the resulting spin dependent Matrix theory scattering amplitudes 
with tree level supergravity \cite{PSW,matrixoneloop} 
does not test any dynamical aspects of the M theory proposal, 
but is solely due to the right amount of supersymmetry in the problem. 

\vspace{6mm}
\noindent
{\bf Acknowledgement}

We wish to thank A. Waldron, who had also guessed \eqn{exp} independently
through an explicit S-Matrix computation, M. Staudacher and A. Tseytlin
for useful comments.



\begin{thebibliography}{99}
%
\bibitem{CH} M. Claudson and M.B. Halpern, {\it Nucl.\ Phys.} {\bf
B250} (1985) 689;\\
R. Flume, {\it Ann.\ Phys.} {\bf 164} (1985) 189,\\
M. Baake, P. Reinicke, and V. Rittenberg, {\it J. Math.\ Phys.}
{\bf 26.} (1985) 1070. 
%
\bibitem{d0} E. Witten, {\it Nucl.\ Phys.} {\bf B460} (1996)
  335, hep-th/9510135.
%
\bibitem{DWHN}
 B. de Wit, J. Hoppe and H. Nicolai, {\it Nucl.\ Phys.} {\bf B305}
[FS23] (1988) 545.
%
\bibitem{BFSS} T.\ Banks, W.\ Fischler, S.H.\ Shenker and L.\
Susskind, {\it Phys. Rev.} {\bf D55} (1997) {5112}, { hep-th/9610043}.
%
\bibitem{treelevel} M.R. Douglas, D. Kabat, P. Pouliot and S.H. Shenker, 
{\it Nucl. Phys.} {\bf B485} (1997) 85, hep-th/9608024. \\
K. Becker, M. Becker, J. Polchinski and A. Tseytlin,
{\it Phys. Rev.} {\bf D56} (1997) 3174, hep-th/9706072\\
Y.~Okawa and T.~Yoneya,
{\it Nucl. Phys.} {\bf B538} (1999) 67, hep-th/9806108.\\
D. Kabat and W. Taylor, {\it Phys. Lett.} {\bf B426} (1998) 297, 
hep-th/9712185.
%
\bibitem{PSW} J.C. Plefka, M. Serone and A.K. Waldron,
{\it Phys. Rev. Lett.} {\bf 81} (1998) 2866, hep-th/9806081;
{\it JHEP} {\bf 11} (1998) 010, hep-th/9809070.
%
\bibitem{R4} R. Helling, J. Plefka, M. Serone and A. Waldron, 
{\it Nucl.\ Phys.} {\bf B559} (1999) 184, hep-th/9905183.
%
\bibitem{PSS98} S. Paban, S. Sethi and M. Stern, {\it Nucl. Phys.} 
{\bf B534} (1998) 137, hep-th/9805018.
%
\bibitem{SSmore}
S. Paban, S. Sethi and M. Stern, {\it JHEP} {\bf 006} 
(1998) 012, hep-th/9806028;\\
S. Sethi and M. Stern, {\it JHEP} {\bf 06} (1999) 004, hep-th/9903049.
%
\bibitem{HKS} S.~Hyun, Y.~Kiem and H.~Shin, {\it Nucl. Phys.} {\bf B}
558 (1999) 349, hep-th/9903022.
%
\bibitem{oneloop} J.F. Morales, C.A. Scrucca and M. Serone, 
{\it Phys. Lett.} {\bf B 417} (1998) 233, hep-th/9709063;
{\it Nucl. Phys.} {\bf B 534} (1998) 223, hep-th/9801183.
%
\bibitem{matrixoneloop}
P. Kraus, {\it Phys. Lett.} {\bf B419} (1998) 73, hep-th/9709199.\\
I.N. McArthur, {\it Nucl. Phys.} {\bf B 534} (1998) 183, hep-th/9806082. \\
M. Barrio, R. Helling and G. Polhemus,
{\it JHEP} {\bf 05} (1998) 012, hep-th/9801189.
%
\end{thebibliography}
\end{document}